\documentclass[conference,a4paper]{IEEEtran}
\usepackage{graphicx}
\ifCLASSOPTIONcompsoc
\usepackage[caption=false,font=normalsize,labelfont=sf,textfont=sf]{subfig}
\else
\usepackage[caption=false,font=footnotesize]{subfig}
\fi
\usepackage[noadjust]{cite}
\usepackage{tabularx}
\usepackage{array}
\usepackage{multirow}
\usepackage{amsfonts}
\usepackage{subfig}

\newtheorem{theorem}{Example}[section]

\begin{document}

\title{A Better Understanding of the Performance\\of Rate-1/2 Binary Turbo Codes\\that Use Odd-Even Interleavers}

\author
{\IEEEauthorblockN{Konstantinos S. Arkoudogiannis, Christos E. Dimakis}
\IEEEauthorblockA{Department of Electrical and Computer Engineering\\
Aristotle University of Thessaloniki\\
Thessaloniki, Greece\\
Email: \{karkoud, dimakis\}@auth.gr}
\and
\IEEEauthorblockN{Konstantinos V. Koutsouvelis}
\IEEEauthorblockA{Hellenic Organization of Telecommunications (OTE)\\
Thessaloniki, Greece\\
Email: koutote@yahoo.gr}
}

\maketitle

\begin{abstract}
The effects of the odd-even constraint---as an interleaver design criterion---on the performance of rate-1/2 binary turbo codes are revisited. According to the current understanding, its adoption is favored because it makes the information bits be uniformly protected, each one by its own parity bit. In this paper, we provide instances that contradict this point of view suggesting for a different explanation of the constraint's behavior, in terms of distance spectrum. 
\end{abstract}

{ \let\thefootnote\relax\footnotetext{K. S. Arkoudogiannis acknowledges the support of the State Scholarships Foundation of Greece (IKY).} }

\section{Introduction}

Since the advent of turbo codes~\cite{Berrou1993} the interleaver structure has been recognized as the key factor in controlling their performance at moderate--high signal-to-noise ratios (SNRs), due to its distance spectrum shaping ability~\cite{Perez1996}. The most notable efforts towards good designs may be S-random~\cite{DolDiv1995}, code-matched~\cite{Feng2002}, high-spread random~\cite{Crozier2000}, and if we also take memory and throughput requirements into consideration, ARP~\cite{Berrou2004} and QPP~\cite{Sun2005} interleavers. The success of these algorithms stems from the fact that they result in permutation patterns satisfying two important conditions: randomness and spread. The first condition helps a turbo code resemble a random one, as suggested by Shannon's second theorem proof, while the second one improves further its weight distribution~\cite[Ch. 16]{ECC2004}. Thus, their combination offers turbo codes good convergence and asymptotic performance.

Especially in the usual case of half-rate ($R=1/2$) binary turbo codes (in which the parity bits are punctured alternately), another design criterion that is considered to ameliorate the bit error rate (BER) performance is the odd-even one, according to which an information bit in odd (even) position before the interleaver must remain in odd (even) position after it. When this constraint is satisfied, the information bits are offered uniform error protection (UEP), meaning that each one of them is accompanied by its corresponding parity bit; otherwise, some information digits will have two parity bits, while others none. The criterion was introduced in~\cite{Barbulescu1994}, where additionally this effect was identified as the reason for the improvement in the performance of a block interleaver. 

Later works on the interleaver design for rate-1/2 turbo encoders apply this criterion to block interleavers, as well as to random-like ones. More specifically, \cite{Ho1998}~advocates its use in S-random and symmetric interleavers, providing simulations for a turbo code with and without the odd-even constraint. A similar comparative study is carried out in~\cite{Hanzo2007} showing its contribution to a random and, even more, to a block interleaver's performance. The authors in~\cite{Yan2009} make use of the odd-even property in order to further improve a quadratic interleaver, while in~\cite{Pajovic2009} an odd-even cyclic shift interleaver is chosen as part of a turbo encoder for underwater communications. Each one of these works justifies the function or the adoption of the criterion by the argumentation phrased previously, implying causality between UEP and better performance. Reference~\cite{Ma2006} is maybe an exception in which a random interleaver is noted to perform worse when meeting the odd-even requirement, but no explanation is given. 

In this paper, we review the impact of the odd-even criterion on rate-1/2 turbo codes, when they employ random, high-spread random and block interleavers. The instances of the (properly punctured) LTE~\cite{3GPP} and Berrou~\cite{Berrou1993} turbo codes are considered. In more detail, we notice that in some cases of the LTE code the simulated results do not support the UEP way of thinking (UEP theory or UEP argument in the text), while in others further explanation is needed. These observations motivate us to rethink the criterion's impact through an alternative approach, that of the resulting distance spectrum. Apart from showing the suitability of this firmly established method for all the examined cases, more importantly we conclude on two properties of the criterion determining the weight distribution and the performance. 

\section{The Random Interleaver Case}

\subsection{Motivation}

Our study begins with the LTE turbo code, the parity bits of which were undergone alternate puncturing for achieving a coderate of 1/2. We measured its performance over the binary-input AWGN channel for medium interleaver lengths, i.e., 512 and 4096 bits, and over the ensemble of random and the ensemble of random odd-even\footnote{The algorithm for the construction of a randomOE interleaver is quite simple: generate at random a number, say \textit{j}, as the candidate interleaved position of the \textit{i}-th bit; if both \textit{i} and \textit{j} are odd (even) store it and do the same for the next bit, else drop it and generate a new number.} (randomOE) interleavers. The reason for simulating code ensembles (i.e., hundreds of thousands of interleavers) was for our study to be more objective, not being possibly dependent on the special structure of a single interleaver. The BER curves we obtained are illustrated in Fig.~\ref{LTEr}. The decoders used the log-MAP algorithm, the number of iterations was 10, and both encoders were terminated as in~\cite{3GPP}. Each SNR point was being simulated until $10^4$ erroneous bits were found.

For a given length, we observe that in the waterfall region the two ensembles have almost the same performance, but what is more surprising is that the random one exhibits a slightly lower error floor. We wish to highlight that one should not deduce that every randomOE interleaver is inferior to any random one; it is the \textit{average} performance of the ensembles that is measured, so the proper interpretation is that the random ensemble contains more good interleavers than the randomOE one does. By combining this clarification with the last observation, we come to realize that there are many instances contradicting the UEP argument. In other words, protecting uniformly the information bits does not always imply better error rates. Since an explanation in UEP terms seems difficult, we suspect that the error floor region paves the way for a clearer understanding of the odd-even constraint's behavior, through the study of the distance spectra.

\subsection{Listening to the Spectra}

Having realized that a convergence analysis via EXIT charts seems---at least in this instance---uninformative, we examined if an asymptotic performance study could answer for the observations. The well-established metric for such a study is the turbo code's free distance and its corresponding multiplicity. These parameters not only determine the performance at high SNRs, but provide reliable enough information about it in the moderate-SNR regime as well.\footnote{The latter is true especially when the distance spectrum is \textit{thin}~\cite{Perez1996}, i.e., the multiplicities of low-weight terms are small, which is a safe assumption for random-like and relatively large interleavers~\cite[Ch. 16]{ECC2004}.} Consequently, we averaged the free-distance terms of a great number of interleavers by exhaustive search over each ensemble, using the algorithm in~\cite{GPB2001}. We opted for this computational method because it is consistent with the aforementioned concept of the evaluation of code \textit{ensembles}. The results are reported in Table I. For each ensemble, we denote by $ \bar{d}_{free} $ the mean of the free distances, by $ \bar{N}_{free} $ their average codeword multiplicity (i.e., the mean number of the minimum-weight codewords), and by $ \bar{w}_{free} $ their average bit multiplicity (i.e., the average total weight of the information sequences leading to each interleaver's minimum-weight codewords). 

There are some observations one can make; first, for a given ensemble and length, the relationship between $ \bar{N}_{free} $ and $ \bar{w}_{free} $ (with the last one being almost twice as great as the first one) reveals how significantly the weight-2 information sequences contribute to the determination of the free distance, hence to the performance of a turbo code at low BERs (say below $10^{-5}$). This has been recorded and studied from the very beginning of turbo codes' history~\cite{Perez1996,Benedetto1996} and more recently as well~\cite{Chatzigeorgiou2009}. Based on this fact, our following study will be concerned only with such sequences. Another thing we notice is that, for a given length, the mean free distance of the random ensemble is slightly larger than the randomOE's one. This provides a partial explanation for the observed error floors. What makes the difference, however, is the average codeword multiplicity; being almost double in the randomOE case seems the main reason for which this ensemble performs---on the average---worse. 
\begin{figure}[!t]
\centering
\includegraphics[width=0.95\linewidth]{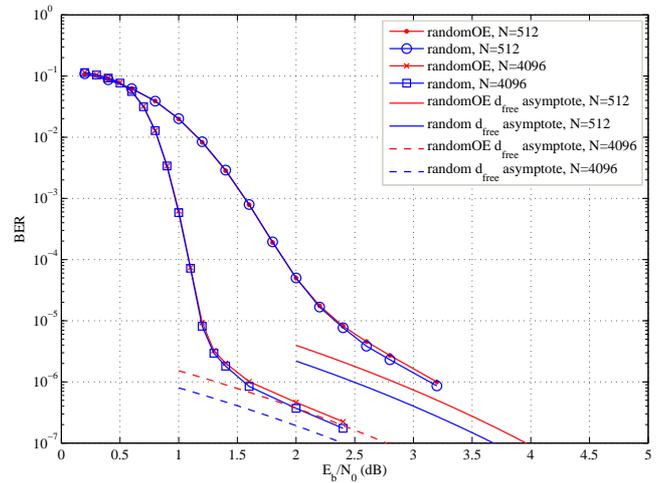}
\caption{BER performance of the LTE random and random odd-even code ensembles, and corresponding average free-distance ML asymptotes. Interleaver lengths: 512 and 4096 bits.}
\label{LTEr}
\end{figure}

In an effort to find the origin of this difference, we kept a log of the information sequences generating for each interleaver the minimum-weight codewords. We discovered that they were mostly those weight-2 self-terminating\footnote{These sequences make the trellis diverge from its zero state and soon remerge with it before its termination, producing a codeword with low weight.} sequences that had their two 1's at distance equal to 7 before as well as after interleaving. A justification can now be attempted by means of probabilities: the 1's of a weight-2 sequence preserve their distance after being permuted with probability roughly $2/N$ for a random interleaver (with the assumption that its size $N$ is much larger than this distance), whereas with $4/N$ for a randomOE one. Therefore, it is expected that such pairings but others too, e.g., distance 7 in the upper encoder and 21 in the lower one, will be happening twice more frequently when the odd-even condition is met.
\begin{table*}[!t]
\renewcommand{\arraystretch}{1.3}
\caption{Mean Values of the Free Distances, Multiplicities and Total Information Weights of the LTE Random and Random Odd-Even Code Ensembles}
\centering
\begin{tabularx}{0.96\textwidth}{ c | c | >{\centering} X >{\centering} X | >{\centering} X >{\centering} X | >{\centering} X X<{\centering}}
\hline
\multirow{2}{*}{ \bfseries Interleaver Length } & \multirow{2}{*}{ \bfseries Number of Interleavers Examined } & \multicolumn{2}{c|}{ $ \bar{d}_{free} $ } & \multicolumn{2}{c|}{ $ \bar{N}_{free} $ } & \multicolumn{2}{c}{ $ \bar{w}_{free} $ }\\
\cline{3-8}
 & & \fontsize{7.5pt}{7.5pt}\textit{\textbf{random}} & \fontsize{7.5pt}{7.5pt}\textit{\textbf{randomOE}} & \fontsize{7.5pt}{7.5pt}\textit{\textbf{random}} & \fontsize{7.5pt}{7.5pt}\textit{\textbf{randomOE}} & \fontsize{7.5pt}{7.5pt}\textit{\textbf{random}} & \fontsize{7.5pt}{7.5pt}\textit{\textbf{randomOE}}\\
\hline
512 & 262\,144 & 7.865 & 7.783 & 2.273 & 3.953 & 5.004 & 8.437\\
4096 & 10\,240 & 8.125 & 8.036 & 2.311 & 4.16 & 4.692 & 8.4\\
\hline
\end{tabularx}
\end{table*}
As these calculations hold regardless of the component convolutional encoders, they admit a more general interpretation: the odd-even constraint \textit{inherently} ``encourages" several unwanted permutations, and, what is more, this is true not only for random, but for random-like (e.g., S-random) interleavers as well, due to the degrees of freedom during the construction of a random or random-like interleaver being halved when the constraint must be additionally satisfied. Consequently, certain low-weight terms of the distance spectrum may acquire an increased multiplicity (like the free-distance terms in our case), limiting the performance of the respective code at moderate or higher SNRs. 

The evidence about $ \bar{N}_{free} $ is indeed enlightening and, along with our comments above, adequate enough for understanding Fig.'s~\ref{LTEr} turbo curves. This is made more illustrative by plotting in the same graph the free-distance maximum-likelihood (ML) asymptotes, as it has already been done, by substitution of the values of Table~I into the following union bound approximation~\cite{Perez1996}:
\begin{equation}
\label{asym}
BER \approx \frac{\bar{w}_{free}}{N} \mathrm{Q} \left( \sqrt{\bar{d}_{free}R\frac{2E_b}{N_0}} \right),
\end{equation}
where $N$ is the interleaver length, $R$ the coderate, $E_b$ the energy per information bit, and $N_0$ the one-sided power spectral density of the AWGN. For a given length, the asymptotes having almost the same slope signifies the articulated difference in the free-distance multiplicities, in the same manner the slopes of the simulated curves intimate. 

Summarizing, the free-distance approach permitted the derivation of a feature of the odd-even criterion unfolding itself under random-like interleaving, namely, the doubling of the probability of some undesirable permutations, which, in many cases, degrades the performance at moderate--high SNRs. The latter could not be explained by the conventional UEP theory. However, this feature is only half the truth; in the next Section we shall describe an opposing property of the same criterion that prevents the distance spectrum from getting denser. 

\section{The High-Spread Random Interleaver Case}

\begin{figure}[!t]
\centering
\subfloat[][]{
\includegraphics[width=0.95\linewidth]{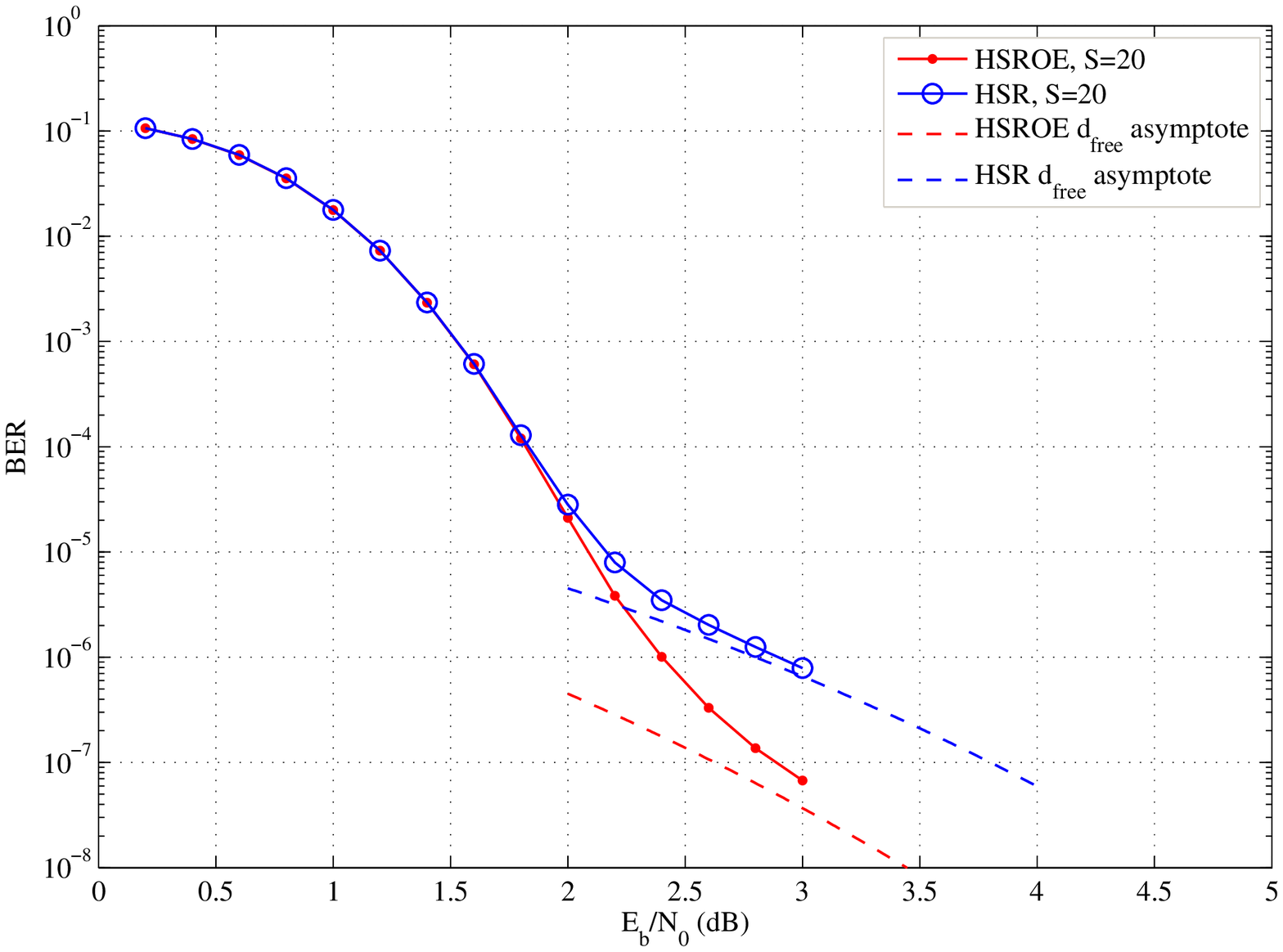}\label{LTEs20}}\\
\subfloat[][]{
\includegraphics[width=0.95\linewidth]{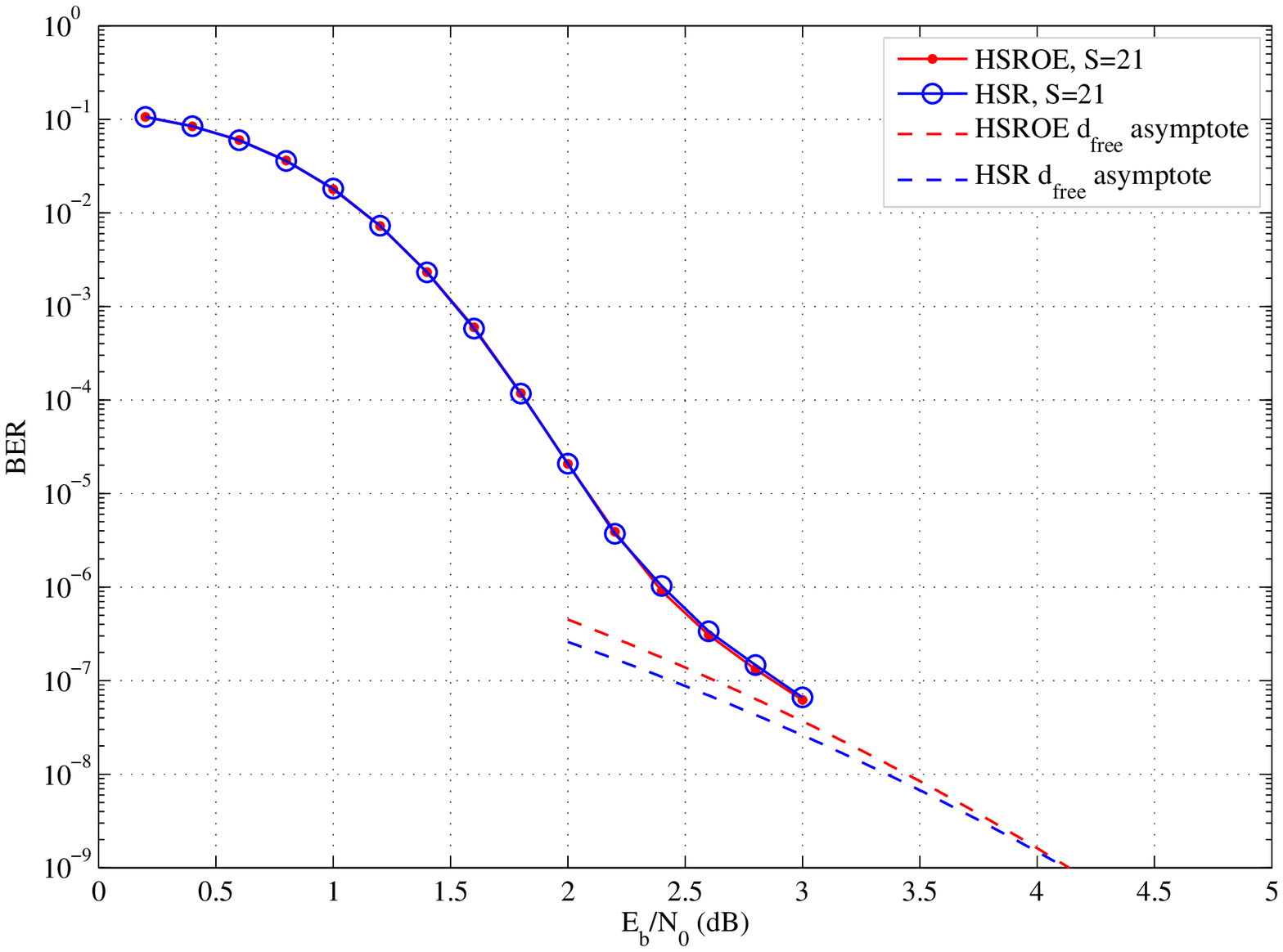}\label{LTEs21}}\\
\caption{BER performance of the LTE high-spread random and hish-spread random odd-even code ensembles, and corresponding average free-distance ML asymptotes. The S-parameter takes the values (a) $S=20$, and (b) $S=21$. Interleaver length: 512 bits.}
\end{figure}

To further illuminate the effects of the odd-even criterion on the weight distribution, we examine the performance of the same encoder for an interleaver size of 512 bits over the high-spread random (HSR) and high-spread randomOE (HSROE) ensembles. We restate here the high-spread criterion~\cite{Crozier2000}: two information bits, say the $i$-th and $j$-th, being at distance less than $S$ must be permuted at distances greater than $S-|i-j|$, that is,
 \begin{displaymath}
|\pi(i)-\pi(j)|>S-|i-j|,\ \ \mathrm{if}\ |i-j|<S,
\end{displaymath}
where $ \pi(.) $ is the permutation function. In order to efficiently simulate the ensembles we took advantage of the algorithm described in~\cite{Koutote2007}. The results are shown in Figs~\ref{LTEs20} and~\ref{LTEs21} with $S$ taking the values 20 and 21, respectively. The simulation details are the same as those in Section II. 

As before, we see that the odd-even criterion has little, if any, effect on the waterfall regions. On the other hand, we notice that in Fig.~\ref{LTEs20} the error floor of the HSROE ensemble is undoubtedly better, which is in agreement with the UEP argument, whereas in Fig.~\ref{LTEs21} the error floors are hardly distinguishable. The latter as compared with the former creates some puzzlement, which---we believe---cannot be resolved in the context of the UEP theory. Thus, as in Section II, we resort to the ensembles' free-distance terms, pursuing a deeper understanding of the way the odd-even criterion behaves. The results are reported in Table II.
\begin{table*}[!t]
\renewcommand{\arraystretch}{1.3}
\caption{Mean Values of the Free Distances, Multiplicities and Total Information Weights of the LTE High-Spread Random and High-Spread Random Odd-Even Code Ensembles}
\centering
\begin{tabularx}{0.96\textwidth}{ c | c | >{\centering} X >{\centering} X | >{\centering} X >{\centering} X | >{\centering} X X<{\centering}}
\hline
\multirow{2}{*}{ \bfseries S-Parameter } & \multirow{2}{*}{ \bfseries Number of Interleavers Examined } & \multicolumn{2}{c|}{ \bfseries $ \bar{d}_{free} $ } & \multicolumn{2}{c|}{ \bfseries $ \bar{N}_{free} $ } & \multicolumn{2}{c}{ \bfseries $ \bar{w}_{free} $ }  \\
\cline{3-8}
 & & \fontsize{7.5pt}{7.5pt}\textit{\textbf{HSR}} & \fontsize{7.5pt}{7.5pt}\textit{\textbf{HSROE}} & \fontsize{7.5pt}{7.5pt}\textit{\textbf{HSR}} & \fontsize{7.5pt}{7.5pt}\textit{\textbf{HSROE}} & \fontsize{7.5pt}{7.5pt}\textit{\textbf{HSR}} & \fontsize{7.5pt}{7.5pt}\textit{\textbf{HSROE}}\\
\hline
20 & 262\,144 & 9 & 11.82 & 13.339 & 13.512 & 26.673 & 27.357\\
21 & 262\,144 & 10.807 & 11.821 & 3.386 & 13.525 & 6.841 & 27.383\\
\hline
\end{tabularx}
\end{table*}

Once again we realize the domination of the weight-2 information sequences. For almost all the interleavers, the free-distance term originated from such sequences, as can been verified by the relationship between $ \bar{N}_{free} $ and $ \bar{w}_{free} $  (for a given ensemble and S-parameter). As for the spectral factors of interest, in the $S=20$ case they inform us sufficiently enough about the superiority of the HSROE ensemble, owing to its higher average free distance. However, in the $S=21$ case we cannot draw such a clear conclusion, due to the contradicting differences between $ \bar{N}_{free} $ and $ \bar{d}_{free} $. Therefore, we turn onto the free-distance ML asymptotes, which have been already plotted in Figs~\ref{LTEs20} and~\ref{LTEs21}. We see that they justify convincingly enough the ensembles' error floors for both values of $S$, illuminating the considerable contribution of the odd-even criterion in the $S=20$ case, and the asymptotic indistinguishability between the two curves in the other case. About Fig.~\ref{LTEs21}, we would like to clarify that the initial slight disagreement between the relative positions of the asymptotes and those of the turbo curves should be apparent, mainly because of the relatively short interleaver size; a more accurate plot of the asymptotes at this region would require more spectral terms (as we are going to see in Section IV for another code). At any rate, we are well informed about the similarity of the curves.

A natural and most important question is why the HSROE ensemble's performance in Fig.~\ref{LTEs20} remains nearly unchanged in Fig.~\ref{LTEs21}, whereas the performances of the HSR ensembles differ so much. The answer would possibly shed more light on the role the odd-even criterion plays in the shaping of the distance spectrum. Having observed that the first term of the spectra almost always comes from weight-2 information sequences, we studied the way this happens and outline it by an example.

\begin{theorem} The component convolutional encoders of the LTE turbo encoder have primitive feedback polynomials ($ 1+D^2+D^3 $) with cycle length\footnote{The cycle length (CL) of a convolutional encoder is defined as the smallest positive integer $ k $ for which its feedback polynomial $ p(D) $ divides $ 1+D^k $ over $ GF(2) $, i.e., $ CL \doteq \min \{k\in \mathbb{N^*}: (1+D^k) \bmod p(D)=0\}$. } equal to 7. This means that for a weight-2 information sequence to be self-terminating, its two 1's, say the $ i $-th and $ j $-th, must be at distances which are integer multiples of 7. The minimum-weight codewords were produced by such information sequences that continued to be self-terminating after interleaving. 

In the $S=20$ case, as far as HSR interleaving is concerned, if $ |i-j| $ equals 7, their images are imposed to lie at a distance greater than $20-7=13$, and, in order for the permuted sequence to remain self-terminating, multiple of 7. So distance 14 is the first trap, to which the two 1's were indeed prone. On the other hand, when the odd-even condition is met, these 1's inescapably remain at odd distance after the interleaver, so as the distance 14 is forbidden, the first trap for them is distance 21, into which they were falling but producing larger free distances. For this reason the HSROE ensemble is better in the error floor region. However, by increasing $S$ by one, a HSR interleaver makes such an information sequence avoid the distance-14 trap, thus produce larger free distances.~~~~~~$\diamond$
\end{theorem}

The above example unveils a new aspect of the odd-even criterion, namely, that of eliminating some ``bad" permutations, hence increasing (possibly) the free distance, and reducing the multiplicities of certain low-weight spectral terms or even suppressing them completely. What is more, the interplay of the criterion with the spread one makes the interleaver work somehow like a code-matched interleaver~\cite{Feng2002}, which is designed so that some specific unwanted permutations are avoided, apart from those ``broken'' by the spread criterion. What we describe is really a countereffect of the odd-even constraint, as opposed to the one described in Section II. Concerning the UEP theory, while it provides a good justification for Fig.'s~\ref{LTEs20} curves, it lacks persuasiveness when comparing them with the ones of Fig.~\ref{LTEs21}.

\section{The Block Interleaver Case}

\begin{figure}[t]
\centering
\includegraphics[width=0.94\linewidth]{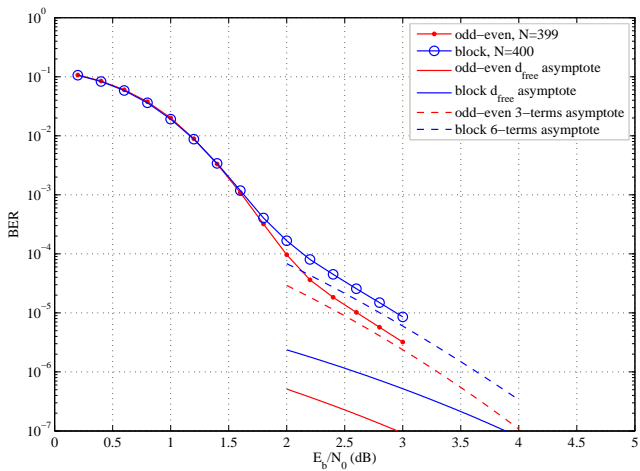}
\caption{BER performance of the Berrou code for the odd-even and block interleavers with lengths 399 and 400 bits, respectively. Corresponding one-term and multiple-terms ML asymptotes.}
\label{Berrou}
\end{figure}

Block interleavers, however useful as channel interleavers for slow-fading environments or as outer interleavers in conventional concatenated coding schemes, do not perform well as a turbo encoder's part---especially the larger lengths---because they cancel the spectral thinning effect~\cite[Ch. 16]{ECC2004}. As a consequence, they suffer from fairly pronounced error floors. Despite this, we are interested in seeing whether a distance spectrum viewpoint can justify the simulated results obtained by Barbulescu and Pietrobon, when they first introduced the odd-even interleaver. For this purpose, we simulated the rate $R=1/2$ Berrou turbo encoder for the same interleaver lengths and structures as in~\cite{Barbulescu1994}. The first permuter was a block one (i.e., writes row-wise, reads column-wise) with size 400 bits (20x20), while the second one was also a block one of size 399 bits, but with odd rows and columns (21x19) which naturally results in an odd-even interleaver. Our simulations are depicted in Fig.~\ref{Berrou}. Their technical details are as in Section II, except for the termination rule that followed~\cite{DivPol1995}. 

Once again the difference is observed mainly in the error floor region. This time, however, our approach was slightly modified: apart from the first term, we also computed some more terms of the distance spectra. The reasons for this change were the relatively short sizes of the interleavers, but predominantly their structures; being highly nonrandom, the multiplicities of low-weight terms were expected to be very large, so a free-distance approach in a sense would be misleading at moderate SNRs. The computation of higher spectral terms stopped when the aforementioned cancellation of the spectral thinning effect was observed. The results are reported in Tables~\ref{table3} and~\ref{table4}, for the odd-even and block interleaver, respectively. 

At first, we see that the odd-even interleaver has larger free distance, but also smaller bit multiplicity. These facts are adequate in their own right to reliably predict the asymptotic superiority of this interleaver, which seems to be supported by the simulations, but do not suffice for such an estimate to be safe in a practically more interesting SNR region. The free-distance ML asymptotes (plotted in the same figure) vividly illustrate this. However, redrawing the asymptotes by taking into account all the listed terms of the corresponding tables [by summing~(\ref{asym}) over each term] successfully addresses this issue (as expected), crediting eventually the distance spectrum approach. Finally, a most significant realization arises by reflecting upon the differences between Tables~\ref{table3} and~\ref{table4}; since the first property of the odd-even criterion (described in Section~II) has been neutralized by the nonrandom structure of the interleavers, the ``elimination property" (referred to in Section~III) may well account for the odd-even interleaver's thinner spectrum.

\section{Conclusion}

The impact of the odd-even constraint on the performance of rate-1/2 binary turbo codes employing random-like and block interleavers has been reexamined. Its operation was considered from a different perspective than the current one, by associating it with the distance spectrum. The underlying motivation was the existence of many instances in which the constraint worsens the performance at low BERs, which in the context of the existing theory are not predicted and seem unexplainable. Instead, the distance spectrum approach has been expectedly sufficient in all the examined cases. More importantly, it lent insight on how the odd-even criterion takes part in the determination of the weight distribution through two antagonistic processes, namely, making some unwanted permutations more probable, while excluding others.

Interesting extensions of this work would be a more thorough examination of how the criterion interacts with spread criteria, as well as how it affects the performance of turbo trellis-coded modulation systems. As a concluding remark, we could say that the odd-even property is primarily a means of distance spectrum shaping, as our study suggests. 

\begin{table}[!t]
\renewcommand{\arraystretch}{1.3}
\caption{First Spectral Terms of the Berrou Turbo Code with the Odd-Even Interleaver}
\label{table3}
\centering
\begin{tabular*}{0.96\columnwidth}{c|c|c|>{\centering}m{1.5 cm}}
\hline
\bfseries Spectral Terms & \bfseries Hamming Weight & \bfseries Multiplicity & \bfseries Total Information Weight
\tabularnewline
\hline
1st & 8 & 1 & 1
\tabularnewline
2nd & 11 & 1 & 1
\tabularnewline
3rd & 12 & 382 & 1523
\tabularnewline
\hline
\end{tabular*}
\end{table}

\begin{table}[!t]
\renewcommand{\arraystretch}{1.3}
\caption{First Spectral Terms of the Berrou Turbo Code with the Block Interleaver}
\label{table4}
\centering
\begin{tabular*}{0.96\columnwidth}{c|c|c|>{\centering}m{1.5 cm}}
\hline
\bfseries Spectral Terms & \bfseries Hamming Weight & \bfseries Multiplicity & \bfseries Total Information Weight
\tabularnewline
\hline
1st & 7 & 1 & 2
\tabularnewline
2nd & 8 & 2 & 3
\tabularnewline
3rd & 9 & 1 & 2
\tabularnewline
4th & 10 & 4 & 8
\tabularnewline
5th & 11 & 4 & 7
\tabularnewline
6th & 12 & 839 & 3350
\tabularnewline
\hline
\end{tabular*}
\end{table}

\bibliographystyle{IEEEtran}
\IEEEtriggeratref{15}
\bibliography{IEEEabrv,pape}

\end{document}